\documentclass[english,aps,prl,showpacs,amsmath,amssymb,twocolumn]{revtex4}
\usepackage[T1]{fontenc}
\usepackage[latin1]{inputenc}
\usepackage{graphicx}

\makeatletter

\usepackage{dcolumn}
\usepackage{bm}
\usepackage{epsf}

\usepackage{babel}
\makeatother
\begin{document}

\title{On the measurement probability of quantum phases}

\author{Thomas Sch\"urmann}

\affiliation{J\"ulich Supercomputing Centre, J\"ulich Research Centre, 52425 J\"ulich, Germany}
%\email{<th.schuermann@gmail.com>}

\begin{abstract}
We consider the probability by which quantum phase measurements of a given precision can be done successfully. The {\it least upper bound} of this probability is derived and the associated optimal state vectors are determined. The probability bound represents an unique and continuous transition between macroscopic and microscopic measurement precisions.
\end{abstract}


\maketitle

The classical picture for the evolution of a single-mode electromagnetic field is simply determined by an amplitude (specifying the strength of the field) and a phase (specifying the zeros of the field). On the other hand, the concept of electromagnetic phase as an observable quantity is a long-standing problem of quantum optics and it has been the question whether there exists a phase observable that is canonically conjugate to the number observable for a single-mode field. The quantum mechanical description of phase was first considered by London \cite{L26} and Dirac \cite{D27}. An obvious way of defining an operator for the phase is by polar decomposition of the photon annihilation operator $\hat{a}=e^{i\hat{\phi}}\sqrt{\hat{N}}$. The phase operator $\hat\phi$ defined in this way is equivalent to that considered by Dirac \cite{D27}, who obtained the commutator $[\hat\phi,\hat{N}]=i$ by employing the correspondence between commutators and classical Poisson bracket. Formally, this would imply the uncertainty relation
\begin{eqnarray}\label{un1}
\sigma_\phi\sigma_N\geq\frac{1}{2}
\end{eqnarray}
with $\sigma_\phi$ and $\sigma_N$ are the standard deviations of $\phi$ and $N$. The difficulties of Dirac's approach were clearly pointed out by Susskind and Glogower \cite{SG64}. Firstly, the relation (\ref{un1}) would imply that a well-defined number state would have a phase standard deviation greater than $2\pi$. This is a consequence of the fact that Dirac's commutator does not take account of the periodic nature of the phase. Furthermore, the exponential operator $e^{i\hat{\phi}}$ derived from this approach is not unitary and thus does not define a Hermitian operator. This is why it is often accepted that a well-behaved Hermitian phase operator does not exist \cite{SG64,CN68}. Therefore, arguments based on the Heisenberg relation (\ref{un1}) cannot hold in general.

Actually, the standard deviation offers a reasonable measure of the spread of values when the distribution in question is of a simple "single hump" type. In particular it is a very good characteristic for a Gaussian distribution since it measures directly the half-width of this distribution. However, when the distribution is not of a simple type (for example, has more than one hump)  the standard deviation loses much of its usefulness as a measure of uncertainty.

The aim of the present contribution is to introduce the probability by which a successful phase measurements of a given precision can be done. The {\it least upper bound} of this probability is determined and the corresponding (optimal) state vectors are computed.

In order to specify phase measurements, the probability distribution for the measurement result can be determined using {\it positive operator-valued measures}. This approach was first considered by Helstrom \cite{H76}, and is also considered in \cite{SSW89,SS91}. Precisely, let $\cal{H}$ be a complex separable Hilbert space, $(|n\rangle)_{n\geq 0}$ an orthonormal basis, and $N=\sum_{n=0}^\infty n|n\rangle\langle n|$ the associated number observable. If the phase density  $F_\phi$ treats all phases equally, it should be invariant under phase translation, $e^{i\theta N}F_\phi e^{-i\theta N}=F_{\phi+\theta}$, generated by the number observable. In this case, the general form of $F_\phi$ is
\begin{eqnarray}\label{M}
F_\phi=\frac{1}{2\pi}\,\sum_{n,m=0}^\infty c_{nm}\,e^{i(n-m)\phi} \;\;|n\rangle\langle m|
\end{eqnarray}
where $(c_{nm})$ is the associated phase matrix. For the integral of the probability to equal 1, we must have
\begin{eqnarray}\label{normierung}
\int_{-\pi}^{\pi}F_\phi\, d\phi =1.
\end{eqnarray}
Applying this to (\ref{M}) above we find that \cite{delta}
\begin{eqnarray}\label{diagonale}
\sum_{n=0}^\infty c_{nn}\;|n\rangle\langle n|=1.
\end{eqnarray}
This means that the diagonal elements $c_{nn}$ must all be equal to 1. The additional condition of positive definite probabilities, together with the above result means that all of the $c_{nm}$ must have absolute values between 0 and 1. In general, real measurements will give smaller values of $c_{nm}$, and the closer these are to 1 the better the phase measurement is. In \cite{LVBP95} it is shown that the additional condition that a number shifter does not alter the phase distribution gives $c_{nm}=1$, corresponding to the canonical measure \cite{LVBP95}
\begin{eqnarray}\label{pom}
E_\phi=\frac{1}{2\pi}\,\sum_{n,m=0}^\infty e^{i(n-m)\phi} \;\;|n\rangle\langle m|.
\end{eqnarray}
An alternative derivation of this result is by using the maximum likelihood approach \cite{SSW89}.  Note that (\ref{pom}) may be expressed by $dE_\phi=|\phi\rangle\langle\phi|\,d\phi$, where
\begin{eqnarray}\label{state}
|\phi\rangle=\frac{1}{\sqrt{2\pi}}\sum_{n=0}^\infty e^{i n \phi}|n\rangle
\end{eqnarray}

With reference to (\ref{pom}), we now define the precision $\Delta\alpha\in[0,2\pi)$ of a phase measurement corresponding to the vicinity $A_{\alpha}=[\alpha-\frac{\Delta\alpha}{2},\alpha+\frac{\Delta\alpha}{2})$ of any value $\alpha\in[-\pi,\pi)$. The probability of a phase measurement with $\phi\in A_\alpha$, made on a state described by a density operator $\hat{\rho}$, is given by
\begin{eqnarray}\label{PA}
P_\alpha(\Delta\alpha)=\mbox{Tr}[\hat{\rho}\,E_\phi(A_\alpha)].
\end{eqnarray}

On the other hand, the probability to measure a photon number $n\in B_k$ is given by
\begin{eqnarray}\label{PB}
P_k(\Delta k)=\mbox{Tr}[\hat\rho\, E_{N}(B_k)]
\end{eqnarray}
where
\begin{eqnarray}\label{EN}
E_{N}(B_k)=\sum_{n\in B_k}\,|n\rangle\langle n|
\end{eqnarray}
is the value of the spectral measure $E_{N}$ on a set $B_k$ of positive integers.
In order to introduce the precision $\Delta k\in\mathbb{N}$ by which the photon number is measured, we note that the photon number is bounded from below. Therefore, we define the right-sided vicinity $B_{k}\subset\mathbb{N}$ of $k$ by $B_{k}=\{k,k+1,...,k+\Delta k\}$.
In this definition, the integer $k$ is the smallest element of $B_{k}$. Alternative definitions are also possible but typically lead to necessary readjustments in certain cases for $k<\Delta k$. However, it will be seen later that our results are not dependent on the particular subscription. For technical purposes we apply the definition of the minimum integer-subscription. In the case of pure states $\hat\rho=|\psi\rangle\langle\psi|$, we obtain the probability
\begin{eqnarray}\label{PN}
P_k(\Delta k)=\sum_{n=k}^{k+\Delta k}|\psi_n|^2
\end{eqnarray}
and $\psi_n=\langle n|\psi\rangle$ is the number-space amplitude of $\psi$.

Now, we consider the case with an initial photon number preparation of a state $\psi$. A single mode is supposed to emerge in a state according to
\begin{eqnarray}\label{reduc}
\psi\to \psi'=\frac{E_{N}(B_k)\psi}{||E_{N}(B_k)\psi||}.
\end{eqnarray}
Afterwards, the number of photons is given with precision $\Delta k$. In this situation the uncertainty principle suggests that the more accurately the number is measured the greater is the perturbation of the phase of the outgoing state. The conditional probability ${\cal P}_{\alpha k}(\Delta \alpha\,|\,\Delta k;\psi)$ to measure phase $\phi\in A_\alpha$, on the state transformed by the initial number measurement, is given by
\begin{eqnarray}\label{prob}
{\cal P}_{\alpha k}(\Delta \alpha\,|\,\Delta k;\psi)=\frac{||E_{\phi}(A_\alpha)E_{N}(B_k)\,\psi||^2}{||E_{N}(B_k)\,\psi||^2}
\end{eqnarray}
We now ask for the {\it least upper bound} $\lambda_0$ of the measurement probability (\ref{prob}) and we end up in a variation problem in Hilbert space with three degrees of freedom. For fixed precisions $\Delta k$ and $\Delta\alpha$ we are searching for the supremum of (\ref{prob}) by variation of the parameters $\alpha$, $k$ and the state vector $\psi$ of the photon. Actually this variation problem is translation and rotation invariant in Hilbert space and we can simply chose $k=0$ and $\alpha=0$ without loss of generality. After all, we have to consider the following expression
\begin{eqnarray}\label{sup}
\lambda_0=\sup\limits_{\psi\in {\cal H}\setminus\{0\}}
{\cal P}_{00}(\Delta \alpha\,|\,\Delta k;\psi)
\end{eqnarray}
and by using (\ref{pom}) and (\ref{EN}) we explicitly obtain to following expression
\begin{eqnarray}\label{expl}
{\cal P}_{00}(\Delta \alpha\,|\,\Delta k;\psi)=
\frac{\int\limits_{-\Delta\alpha/2}^{\Delta\alpha/2} \left|\frac{1}{\sqrt{2\pi}}\sum\limits_{n=0}^{\Delta k}\psi_n\,e^{in\phi}\right|^2 d\phi}{\sum\limits_{n'=0}^{\Delta k}|\psi_{n'}|^2}.
\end{eqnarray}
Applying the Cauchy-Bunyakovsky inequality we obtain the following general upper bound
\begin{eqnarray}\label{bound}
{\cal P}_{\alpha,k}(\Delta \alpha\,|\,\Delta k;\psi)\leq\frac{\Delta\alpha (\Delta k+1)}{2\pi}
\end{eqnarray}
for every $\Delta\alpha\in[-\pi,\pi)$ and integer $\Delta k\geq 0$. In fig. 1, the set of impossible measurement processes is expressed by the grey shaded triangle.
\begin{figure}[ht]
\includegraphics[width=9.0cm,height=7.0cm]{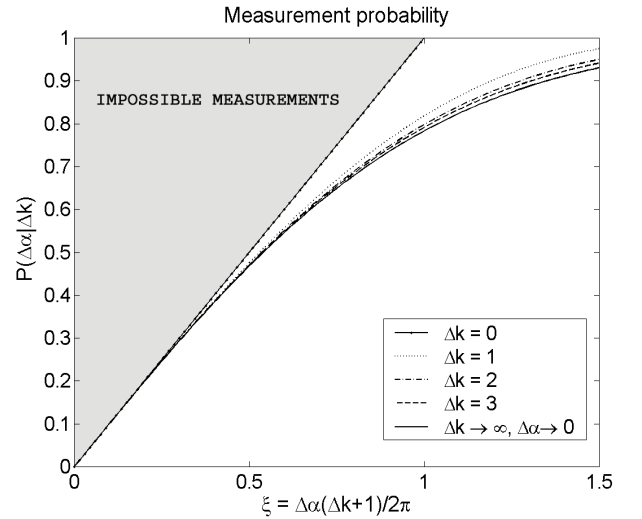}
\caption{Maximum eigenvalue $\lambda_0$ for $\Delta k=0,1,2,3,\infty$ (from left to right). Phase measuring processes with probabilities larger than $\lambda_0$ do not exist.}\label{fig1}
\end{figure}
In order to reach even tighter bounds we explicitly computed the integral in (\ref{expl}) and applied certain trigonometric identities to obtain the following expression
\begin{eqnarray}\label{bi}
{\cal P}_{00}(\Delta \alpha\,|\,\Delta k;\psi)=
\sum\limits_{n,m=0}^{\infty}G_{nm}\,\psi_n\,\psi_m^*
\end{eqnarray}
with normalization condition $\sum_{n=0}^{\Delta k}|\psi_n|^2=1$ and kernel \cite{delta}
\begin{eqnarray}\label{G}
G_{nm}=\frac{1}{\pi} \frac{\sin \frac{\Delta\alpha}{2}(n-m)}{n-m}
\end{eqnarray}
for $0\leq n,m\leq \Delta k$, $0$ otherwise. Obviously, $G$ is self-adjoint and positive definite.
According to (\ref{sup}) and (\ref{bi}), the least upper bound is given by the operator norm of $G$, i.e.
\begin{eqnarray}\label{opnorm}
\lambda_0=||G||,
\end{eqnarray}
and this norm is identical to the largest eigenvalue of $G$. In order to obtain the eigenvalues of $G$ we have to solve the following linear equation for $n=0,1,...,\Delta k$
\begin{eqnarray}\label{eigen}
\sum_{m=0}^{\Delta k}\,G_{nm}\;\psi_m^{(s)}=\lambda_s\;\psi_n^{(s)}
\end{eqnarray}
for $s=0,1,...,\Delta k$. This type of eigenvalue problem has been extensively discussed in \cite{S78} (see also references therein). All eigenvalues are distinct, positive and may be ordered so that $1>\lambda_0>\lambda_1>...>\lambda_{\Delta k}$. In the non-trivial case of $\Delta k>0$ we computed $\lambda_0$ numerically. For increasing values of $\Delta k$, the corresponding bounds approach very fast to the asymptotic case $\Delta k\to\infty$, see fig. 1 (right most continuous line). For the computation of the asymptotic case we introduced the equidistant decomposition $q_m=\frac{m}{\Delta k+1}$, $m=0,1,...,\Delta k$ with increment $\delta q_m=q_{m+1}-q_m$. After substitution into (\ref{eigen}) and a few algebraic manipulations, the discrete eigenvalue problem approaches to the following homogeneous Fredholm integral equation of the first kind
\begin{eqnarray} \label{eig}
\frac{1}{\pi}\,\int_{-1}^{1}\frac{\sin(\frac{\pi}{2}\xi(z-z'))}{z-z'}\;\varphi^{(\nu)}(z')\,d z'
=\tilde\lambda_\nu(\xi)\,\varphi^{(\nu)}(z)\nonumber\\
\end{eqnarray}
with $|z|\leq 1$ and
\begin{eqnarray} \label{xi}
\xi=\frac{\Delta\alpha(\Delta k+1)}{2\pi}.
\end{eqnarray}
From standard theory we know that there are solutions in $L^2([-1,1])$ only for a discrete set of eigenvalues, say $\tilde\lambda_0\geq\tilde\lambda_1\geq...$ and that as $\nu\to\infty$, $\tilde\lambda_\nu\to 0$. It should be noted that the eigenvalues explicitly depend on the parameter $\xi$ and corresponding to each eigenvalue there is a unique (up to normalization) solution $\varphi^{(\nu)}(z)= S_{0\nu}(\pi\xi/2,z)$ called {\it angular prolate spheroidal wave function} \cite{S78,AS}. They are continuous functions of $\xi$ for $\xi\geq 0$, and are orthogonal in $(-1,1)$. Moreover, they are complete in $L^2([-1,1])$. The corresponding eigenvalues are related to a second set of functions called {\it radial prolate spheroidal functions}, which differ from the angular functions only by a real scale factor \cite{S78}. Applying the notation of \cite{S78,AS}, these eigenvalues are
\begin{eqnarray} \label{values}
\tilde\lambda_\nu(\xi)=\xi\,\left[R^{(1)}_{0\nu}(\pi\xi/2,1)\right]^{\,2}
\end{eqnarray}
with $\nu=0,1,2,...$ The properties of the discrete eigenvalue spectrum is discussed in \cite{L65}. Here, we are mainly interested in the properties of the largest eigenvalue $\tilde\lambda_0(\xi)$. It is monotonically increasing and approaches $1$ exponentially in $\xi$. For small values of $\xi$ there is the asymptotic behavior $\tilde\lambda_0(\xi)\sim\xi$.

\newpage{}

\begin{thebibliography}{10}
\bibitem{delta} It is understood here, that when $n=m$ the expression $\frac{\sin(\frac{\Delta\alpha}{2}\,(n-m))}{n-m}$ has the value $\frac{\Delta\alpha}{2}$.
\end{thebibliography}
\end{document}